\documentclass[12pt]{iopart}
\usepackage{cite}
\usepackage[dvipdfmx]{graphicx}
\usepackage{color}
\begin{document}

\title[An arm length stabilization system for KAGRA and future GW detectors]{An arm length stabilization system for KAGRA and future gravitational-wave detectors}
\author{
T.~Akutsu$^{1, 2}$, 
M.~Ando$^{3, 4, 1}$, 
K.~Arai$^{5}$, 
K.~Arai$^{6}$, 
Y.~Arai$^{6}$, 
S.~Araki$^{7}$, 
A.~Araya$^{8}$, 
N.~Aritomi$^{3}$, 
Y.~Aso$^{9, 10}$, 
S.~Bae$^{11}$
% ...
}

\address{${}^{1}$ Gravitational Wave Project Office, National Astronomical Observatory of Japan (NAOJ), Mitaka City, Tokyo 181-8588, Japan}
\address{${}^{2}$ Advanced Technology Center, National Astronomical Observatory of Japan (NAOJ), Japan}
\address{${}^{3}$ Department of Physics, The University of Tokyo, Bunkyo-ku, Tokyo 113-0033, Japan}
\address{${}^{4}$ Research Center for the Early Universe (RESCEU), The University of Tokyo, Bunkyo-ku, Tokyo 113-0033, Japan}
\address{${}^{5}$ California Institute of Technology, Pasadena, CA 91125, USA}
\address{${}^{6}$ Institute for Cosmic Ray Research (ICRR), KAGRA Observatory, The University of Tokyo, Kashiwa City, Chiba 277-8582, Japan}
\address{${}^{7}$ Accelerator Laboratory, High Energy Accelerator Research Organization (KEK), Tsukuba City, Ibaraki 305-0801, Japan}
\address{${}^{8}$ Earthquake Research Institute, The University of Tokyo, Bunkyo-ku, Tokyo 113-0032, Japan}
\address{${}^{9}$ Kamioka Branch, National Astronomical Observatory of Japan (NAOJ), Kamioka-cho, Hida City, Gifu 506-1205, Japan}
\address{${}^{10}$ The Graduate University for Advanced Studies (SOKENDAI), Mitaka City, Tokyo 181-8588, Japan}
\address{${}^{11}$ Korea Institute of Science and Technology Information (KISTI), Yuseong-gu, Daejeon 34141, Korea}
\address{${}^{12}$ National Institute for Mathematical Sciences, Daejeon 34047, Korea}
\address{${}^{13}$ Department of Earth and Space Science, Graduate School of Science, Osaka University, Toyonaka City, Osaka 560-0043, Japan}
\address{${}^{14}$ School of High Energy Accelerator Science, The Graduate University for Advanced Studies (SOKENDAI), Tsukuba City, Ibaraki 305-0801, Japan}
\address{${}^{15}$ Department of Applied Physics, Fukuoka University, Jonan, Fukuoka City, Fukuoka 814-0180, Japan}
\address{${}^{16}$ Department of Physics, Tamkang University, Danshui Dist., New Taipei City 25137, Taiwan}
\address{${}^{17}$ Department of Physics and Institute of Astronomy, National Tsing Hua University, Hsinchu 30013, Taiwan}
\address{${}^{18}$ Department of Physics, Center for High Energy and High Field Physics, National Central University, Zhongli District, Taoyuan City 32001, Taiwan}
\address{${}^{19}$ Institute of Physics, Academia Sinica, Nankang, Taipei 11529, Taiwan}
\address{${}^{20}$ Department of Physics, University of Toyama, Toyama City, Toyama 930-8555, Japan}
\address{${}^{21}$ Univ.~Grenoble Alpes, Laboratoire d'Annecy de Physique des Particules (LAPP), Universit\'e Savoie Mont Blanc, CNRS/IN2P3, F-74941 Annecy, France}
\address{${}^{22}$ Department of Astronomy, The University of Tokyo, Mitaka City, Tokyo 181-8588, Japan}
\address{${}^{23}$ State Key Laboratory of Magnetic Resonance and Atomic and Molecular Physics, Wuhan Institute of Physics and Mathematics (WIPM), Chinese Academy of Sciences, Xiaohongshan, Wuhan 430071, China}
\address{${}^{24}$ Applied Research Laboratory, High Energy Accelerator Research Organization (KEK), Tsukuba City, Ibaraki 305-0801, Japan}
\address{${}^{25}$ Institute for Cosmic Ray Research (ICRR), KAGRA Observatory, The University of Tokyo, Kamioka-cho, Hida City, Gifu 506-1205, Japan}
\address{${}^{26}$ College of Industrial Technology, Nihon University, Narashino City, Chiba 275-8575, Japan}
\address{${}^{27}$ Graduate School of Science and Technology, Niigata University, Nishi-ku, Niigata City, Niigata 950-2181, Japan}
\address{${}^{28}$ Department of Physics, National Taiwan Normal University, Taipei 116, Taiwan}
\address{${}^{29}$ Institute for Cosmic Ray Research (ICRR), Research Center for Cosmic Neutrinos (RCCN), The University of Tokyo, Kashiwa City, Chiba 277-8582, Japan}
\address{${}^{30}$ Kavli Institute for Astronomy and Astrophysics, Peking University, China}
\address{${}^{31}$ Yukawa Institute for Theoretical Physics (YITP), Kyoto University, Sakyou-ku, Kyoto City, Kyoto 606-8502, Japan}
\address{${}^{32}$ Department of Physics, Graduate School of Science, Osaka City University, Sumiyoshi-ku, Osaka City, Osaka 558-8585, Japan}
\address{${}^{33}$ Nambu Yoichiro Institute of Theoretical and Experimental Physics (NITEP), Osaka City University, Sumiyoshi-ku, Osaka City, Osaka 558-8585, Japan}
\address{${}^{34}$ Institute of Space and Astronautical Science (JAXA), Chuo-ku, Sagamihara City, Kanagawa 252-0222, Japan}
\address{${}^{35}$ Department of Physics, School of Natural Science, Ulsan National Institute of Science and Technology (UNIST), Ulsan 44919, Korea}
\address{${}^{36}$ Institute for Cosmic Ray Research (ICRR), The University of Tokyo, Kashiwa City, Chiba 277-8582, Japan}
\address{${}^{37}$ Graduate School of Science and Technology, Tokyo Institute of Technology, Meguro-ku, Tokyo 152-8551, Japan}
\address{${}^{38}$ Department of Physics, Ewha Womans University, Seodaemun-gu, Seoul 03760, Korea}
\address{${}^{39}$ Department of Computer Simulation, Inje University, Gimhae, Gyeongsangnam-do 50834, Korea}
\address{${}^{40}$ Department of Physical Science, Hiroshima University, Higashihiroshima City, Hiroshima 903-0213, Japan}
\address{${}^{41}$ Institute for Cosmic Ray Research (ICRR), Research Center for Cosmic Neutrinos (RCCN), The University of Tokyo, Kamioka-cho, Hida City, Gifu 506-1205, Japan}
\address{${}^{42}$ Institute for Advanced Research, Nagoya University, Furocho, Chikusa-ku, Nagoya City, Aichi 464-8602, Japan}
\address{${}^{43}$ Department of Physics, Hanyang University, Seoul 133-791, Korea}
\address{${}^{44}$ National Center for High-performance computing, National Applied Research Laboratories, Hsinchu Science Park, Hsinchu City 30076, Taiwan}
\address{${}^{45}$ Institute for Photon Science and Technology, The University of Tokyo, Bunkyo-ku, Tokyo 113-8656, Japan}
\address{${}^{46}$ Institute for Laser Science, University of Electro-Communications, Chofu City, Tokyo 182-8585, Japan}
\address{${}^{47}$ The Applied Electromagnetic Research Institute, National Institute of Information and Communications Technology (NICT), Koganei City, Tokyo 184-8795, Japan}
\address{${}^{48}$ Faculty of Law, Ryukoku University, Fushimi-ku, Kyoto City, Kyoto 612-8577, Japan}
\address{${}^{49}$ Department of Physics, Kyoto University, Sakyou-ku, Kyoto City, Kyoto 606-8502, Japan}
\address{${}^{50}$ School of Optical Electrical and Computer Engineering, The University of Shanghai for Science and Technology, China}
\address{${}^{51}$ Faculty of Engineering, Niigata University, Nishi-ku, Niigata City, Niigata 950-2181, Japan}
\address{${}^{52}$ Center for advanced photonics, Research Institute of Physics and Chemstry (RIKEN), Wako City, Saitama 351-0198, Japan}
\address{${}^{53}$ Optical instrument developement team, Korea Basic Science Institute, Korea}
\address{${}^{54}$ Department of Engineering, University of Sannio, Benevento 82100, Italy}
\address{${}^{55}$ Faculty of Arts and Science, Kyushu University, Nishi-ku, Fukuoka City, Fukuoka 819-0395, Japan}
\address{${}^{56}$ Department of Electronic Control Engineering, National Institute of Technology, Nagaoka College, Nagaoka City, Niigata 940-8532, Japan}
\address{${}^{57}$ Graduate School of Science and Engineering, Hosei University, Koganei City, Tokyo 184-8584, Japan}
\address{${}^{58}$ Faculty of Science, Toho University, Funabashi City, Chiba 274-8510, Japan}
\address{${}^{59}$ Faculty of Information Science and Technology, Osaka Institute of Technology, Hirakata City, Osaka 573-0196, Japan}
\address{${}^{60}$ Department of Information \& Management  Systems Engineering, Nagaoka University of Technology, Nagaoka City, Niigata 940-2188, Japan}
\address{${}^{61}$ National Metrology Institute of Japan, National Institute of Advanced Industrial Science and Technology, Tsukuba City, Ibaraki 305-8568, Japan}
\address{${}^{62}$ University of Camerino, Italy}
\address{${}^{63}$ Istituto Nazionale di Fisica Nucleare, University of Perugia, Perugia 06123, Italy}
\address{${}^{64}$ Faculty of Science, Department of Physics, The Chinese University of Hong Kong, Shatin, N.T., Hong Kong, Hong Kong}
\address{${}^{65}$ Faculty of Science, Niigata University, Nishi-ku, Niigata City, Niigata 950-2181, Japan}
\address{${}^{66}$ Department of Communications, National Defense Academy of Japan, Yokosuka City, Kanagawa 239-8686, Japan}
\address{${}^{67}$ Department of Physics, University of Florida, Gainesville, FL 32611, USA}
\address{${}^{68}$ Department of Physics and Astronomy, Sejong University, Gwangjin-gu, Seoul 143-747, Korea}
\address{${}^{69}$ Department of Astronomy, Beijing Normal University, Beijing 100875, China}


\ead{yenomoto@icrr.u-tokyo.ac.jp}
\vspace{10pt}
\begin{indented}
\item[]\today
\end{indented}

\begin{abstract}
Modern ground-based gravitational wave (GW) detectors require a complex interferometer configuration with multiple coupled optical cavities. 
Since achieving the resonances of the arm cavities is the most challenging among the lock acquisition processes, the scheme called arm length stabilization (ALS) had been employed for lock acquisition of the arm cavities. We designed a new type of the ALS, which is compatible with the interferometers having long arms like the next generation GW detectors. The features of the new ALS are that the control configuration is simpler than those of previous ones and that it is not necessary to lay optical fibers for the ALS along the kilometer-long arms of the detector. 
Along with simulations of its noise performance, an experimental test of the new ALS was performed utilizing a single arm cavity of KAGRA. This paper presents the first results of the test where we demonstrated that lock acquisition of the arm cavity was achieved using the new ALS. We also demonstrated that the root mean square of residual noise was measured to be $8.2\,\mathrm{Hz}$ in units of frequency, which is smaller than the linewidth of the arm cavity and thus low enough to lock the full interferometer of KAGRA in a repeatable and reliable manner.
\end{abstract}

%
\vspace{2pc}

\submitto{\CQG}
%
%
%

\section{Introduction}
The first direct detection of gravitational waves (GWs) by the two LIGO detectors from a binary black hole merger in 2015 \cite{Abbott2016} marked the beginning of an era of GW astronomy. Moreover, the detection of GWs from a binary neutron star merger by the three LIGO-Virgo detectors \cite{Abbott2017a}, and many other electromagnetic follow-up observations \cite{Abbott2017b} provided new insights into compact stars and their mergers. These observations of GW events have proved the importance of direct detection of GWs and detection with multiple GW detectors.
It is planned that KAGRA joins the global GW detector network in late 2019 as the fourth detector \cite{Scenario2018}. It is expected that four-detector-observation enables us to improve the sky coverage of the localizable sources of GWs \cite{wen2010,Scenario2018}, to increase the network duty cycle, and to disentangle polarizations of GWs \cite{takeda2018}.
KAGRA has two unique features that the detector is built underground \cite{Akutsu2018,Somiya2012} and the key components, test masses that respond to GWs, are cooled down to cryogenic temperature \cite{akutsu2019,Somiya2012}. These features will be of great importance for the next generation detectors \cite{Hild2011,abbott2017}.
Therefore intensive development of KAGRA is necessary not only for maximizing the science output of the four-detector network but also for paving the way towards the realization of the next generation detectors.

Currently all the GW detectors at sensitivities high enough to detect GWs are terrestrial laser interferometers. 
Although the working principle of such detectors is as simple as that of the Michelson interferometer, the real optical configurations of them are rather complex. In fact, the designed configurations of the main interferometers of Advanced LIGO, Advanced Virgo, and KAGRA employ multiple optical cavities. To properly operate the detector, all the cavities have to be controlled at their resonances. Thus the control of the main interferometer is essential. Furthermore, it is necessary to have a tractable scheme to achieve the resonances of the cavities, since it is inevitable for the interferometer to lose its lock by external disturbances such as earthquakes. 
The process where the main interferometer cavities are brought to their resonances is called \textit{lock acquisition}.
However, the lock acquisition process is generally not straightforward, due to highly non-linear and cross-coupled response of the optical cavities to change in the cavity lengths \cite{acernese2006,evans2002}. In particular, achieving the resonances of the arm cavities is most challenging because kilometer-long arm cavities come at a cost of very narrow frequency linewidth.

In order to achieve reliable and repeatable lock acquisition, a scheme called arm length stabilization (ALS) has been proposed \cite{mullavey2012}, demonstrated \cite{mullavey2012,Izumi2012} and implemented in Advanced LIGO \cite{Staley2014}. A similar scheme will be installed also in Advanced Virgo \cite{Acernese2015}. The scheme employs two auxiliary lasers that are phase-locked to the main laser of the interferometer, and the second harmonic generation (SHG) light of each auxiliary laser is used to control and stabilize the length of each arm cavity. With the ALS scheme, each cavity length can be kept at an arbitrary point with respect to the resonant points of the cavity. This makes it easy to acquire lock of the central part of the interferometer. Otherwise, control of the central part would be significantly disturbed or even broken due to the arm cavities stochastically passing across resonances. The SHG is used because use of different wavelength from that of the main laser enables us to utilize dichroic coating of mirrors. The finesse of the arm cavities are designed to be so low that achieving the resonance for the SHG light is easy. In addition, it is known that excess frequency noise introduced by the SHG is low enough \cite{yeaton-massey2012}.

KAGRA also adopts ALS as a part of the lock acquisition scheme. KAGRA uses two auxiliary lasers similarly, but the control scheme is different as it is simplified from that of Advanced LIGO in a few aspects. 
In addition, the designed linewidth of the arm cavities is smaller than that of Advanced LIGO or Advanced Virgo, which may make lock acquisition of the KAGRA interferometer even more challenging. Therefore the detailed design and an experimental demonstration of the ALS scheme are crucial for the operation of KAGRA.
In this paper, we report the control scheme of the ALS system of KAGRA, the design of the noise performance, and the results of the performance test utilizing a single arm cavity of KAGRA, where we successfully locked the arm cavity with the scheme.

\begin{figure}
\centering
\includegraphics[width=16cm,clip]{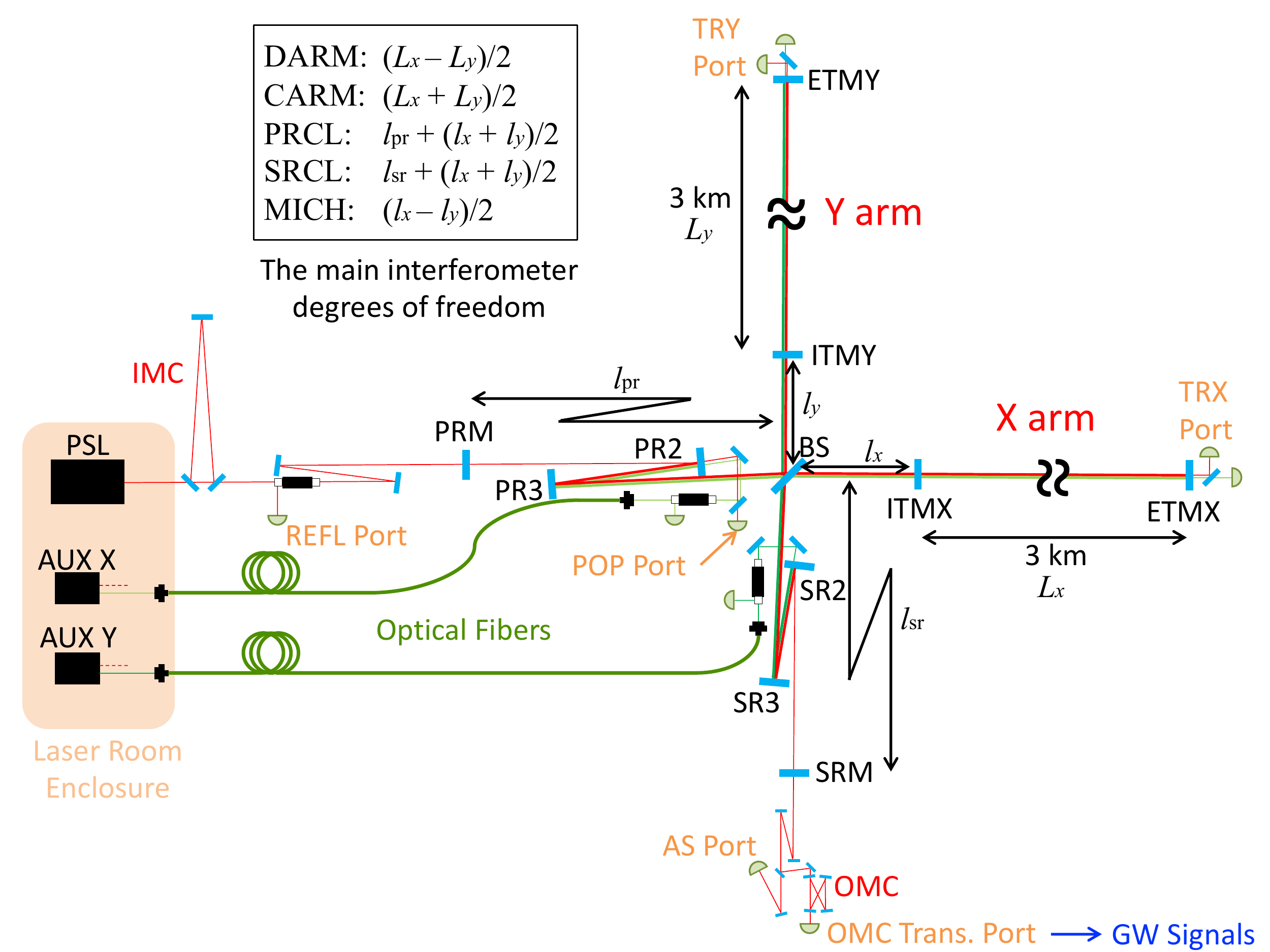}
\caption{Schematic view of the optical configuration of KAGRA. A BS is placed at the center of the interferometer and separate the main laser beam into two arms, namely the X and Y arms. Each arm has a $3\,\mathrm{km}$ long cavity. Interference at the BS is controlled so that almost all the laser power is reflected back to the PRM. The detection ports for the control of the interferometer are also shown. Two dichroic auxiliary laser are placed in the laser room. Their green laser outputs are transferred to the center of the interferometer via optical fibers, and then injected to the main interferometer from the back side of PR2 or SR2. Refer to the main body for the detailed description. The definitions of the five length degrees of freedom of the main interferometer of KAGRA are also described.}\label{config}
\end{figure}
\section{Interferometer of KAGRA}
Figure \ref{config} shows a schematic view of the optical configuration of KAGRA. The optical configuration described here is similar to those of Advanced LIGO and Advanced Virgo, except that four test masses are cooled down to cryogenic temperature. The pre-stabilized laser (PSL) at a wavelength of $1064\,\mathrm{nm}$ is sent to a triangular optical cavity called the input mode cleaner (IMC). The transmitted light subsequently passes through a Faraday isolator in which the reflected beam from the main interferometer is extracted. The main interferometer is composed of four identical test masses (ETMX, ITMX, ETMY, and ITMY), a beam splitter (BS), three power recycling mirrors (PRM, PR2, and PR3), and three signal recycling mirrors (SRM, SR2, and SR3). Each pair of test masses, ETM and ITM, forms an arm cavity \cite{drever1991}. The two arm cavities are called the X and Y arms. The arm cavities and the BS form a Fabry--Perot Michelson interferometer (FPMI). The power recycling mirrors and both ITMs form a cavity called power recycling cavity, where the PRM resonantly reflects back the reflection from the FPMI to build up the laser power at the BS \cite{meers1988}. Similarly, the signal recycling mirrors and both ITMs form another cavity called signal recycling cavity, where the SRM resonantly reflects back the transmission of the FPMI to change the detector response to GW signals so as to optimize the sensitivity of the detector \cite{meers1988,mizuno1993}. The transmission of the SRM, the output of the main interferometer, propagates through a bowtie cavity called the output mode cleaner (OMC). Variations in the optical power in transmission of the OMC corresponds to GW signals \cite{fricke2012}.

There are five length degrees of freedom in the main interferometer. The names and definitions of the degrees of freedom are summarized also in Figure \ref{config}. By utilizing phase modulations applied to the main laser in the laser room \cite{regehr1995}, the information of displacement in each length degree of freedom is obtained at various detection ports \cite{martynov2016,Aso2013}. Error signal for CARM is obtained at the REFL port while error signal for DARM is obtained either at the AS port or the OMC transmission (trans.) port. CARM is sensitive to the frequency of the PSL as the resonant frequency of CARM is much more stable than the PSL frequency in general. Thus the control of CARM is implemented by feeding the CARM signals back to the PSL frequency. In contrast, DARM is much less sensitive to fluctuations of the PSL frequency. So for the reason, the control signal is sent to the ETMs.

For the purpose of lock acquisition of the arm cavities, we have two auxiliary lasers in addition to the PSL in the laser room. 
Each auxiliary laser provides a pair of laser outputs: infrared and green. The infrared beam is the primary output and has a wavelength of $1064\,\mathrm{nm}$. The secondary one is the green laser which is frequency-doubled light of the primary beam, and thus has the wavelength of $532\,\mathrm{nm}$.
The two green lasers are sent to the optical tables in the vicinity of PR2 or SR2 via optical fibers with the length of about $60\,\mathrm{m}$. The green lasers are then injected to the main interferometer from the back side of PR2 or SR2. 
Because the PR2, SR2, and BS are dichroic and transmissive at $532 \mathrm{nm}$, the green lasers injected from PR2 and SR2 are incident to the X and Y arms, respectively.
The optical paths of the green lasers are slightly separated from that of the main laser owing to the wedged substrate and the dispersion of the BS and the ITMs.

\section{Lock acquisition and ALS of KAGRA}
\subsection{Lock acquisition scheme}\label{locksubsection}
Our lock acquisition scheme is similar to that of Advanced LIGO \cite{Staley2014}. It is divided into three steps as follows: (1) lock both arm cavities with the green lasers and keep the arm cavity lengths at off-resonant points for the main laser by the ALS system, (2) lock the vertex interferometer (i.e. PRCL, SRCL, and MICH) and keep them locked using a combination of the interferometric signals that are less sensitive to the carrier field of the main laser e.g., third harmonic demodulation signals \cite{arai2000} or beat note signals using sideband fields non-resonant in the main interferometer \cite{Aso2013,Yamamoto2019}, and finally (3) bring the arm cavity lengths to their resonances and switch the error signals of the arm length degrees of freedom to that obtained by using the carrier field of the main laser. In the step (3), the main laser frequency needs to be controlled with respect to the resonant frequency of the arm cavities with a precision better than the linewidths of the arm cavities. Otherwise, the control of the three degrees of freedom of the vertex interferometer would be disturbed.

\subsection{KAGRA-type ALS}
\begin{figure}
\centering
\includegraphics[width=16cm,clip]{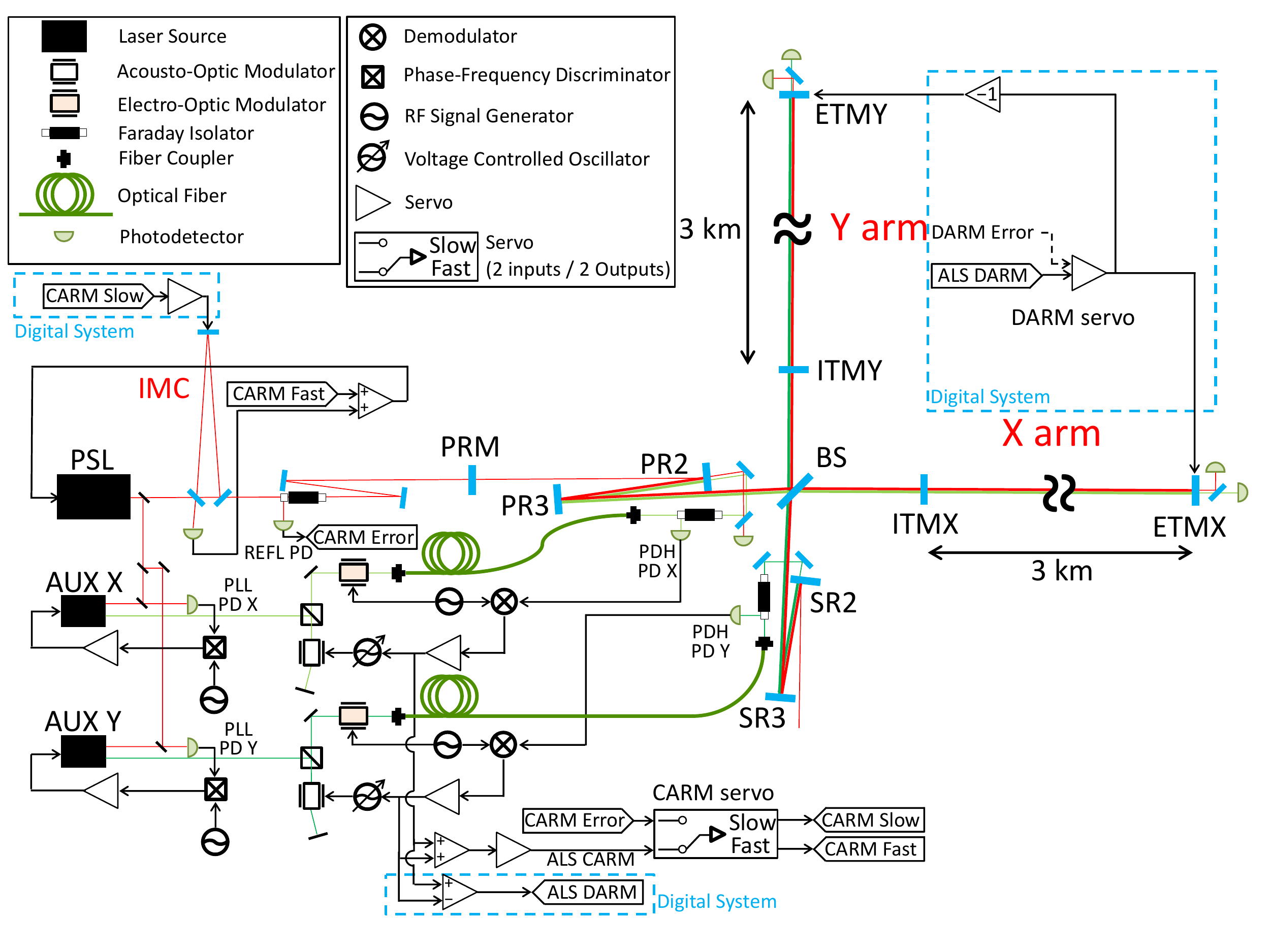}
\caption{Overview of the arm length stabilization system of KAGRA. The frequencies of the two auxiliary lasers are phase-locked to that of PSL. The frequency of the each green laser is controlled by the combination of a double-path acousto-optic modulator (AOM) and a voltage controlled oscillator (VCO) so that the green laser resonates in the corresponding arm cavity. The sum of the two control signals applied to the VCOs is used as an error signal of CARM (ALS CARM), and the difference of the two is used as an error signal of DARM (ALS DARM).}\label{alsconfig}
\end{figure}
The ALS of KAGRA is different from that of Advanced LIGO in their concepts, designs, and control strategy. 
Table \ref{compare} summarizes the main differences between the ALS systems of KAGRA and Advanced LIGO.
The features of the KAGRA ALS can be noted in the following. 
The first feature is that two green beams are injected from the central area, as opposed to the end stations. Thanks to this feature, the necessary length of the optical fibers is not on the same order of the arm length but approximately $60\,\mathrm{m}$, which makes it scalable to those with longer arm length such as the third generation detectors \cite{Hild2011,abbott2017}; phase noise and optical loss associated with the fiber do not increase as the arm length gets long. 
The second is that the configuration is simple in two aspects: less number of the sensors and less number of SHG setups. One possible drawback coming from this feature is that phase noise the lasers pick up as they propagate through the optical fibers directly becomes sensing noise of the ALS system. However, this issue is not critical because fiber phase noise is not too large since the fiber length is short; the noise design and characterization revealed that the ALS system could satisfy the noise requirements, as described in the following sections. Additionally, the fiber noise cancellation technique \cite{ye2003} can be implemented independently if needed, though it would indeed add complexity to the system. We can expect the suppression of a factor of more than $10^3$ below $100\,\mathrm{Hz}$ by setting the cancellation bandwidth to around $10\,\mathrm{kHz}$. 
The third feature is that the ALS CARM and ALS DARM signal are produced by summing the two signals from two arms in electronics or real-time controllers, respectively. This increases the flexibility of the ALS control loops. 
\begin{table}
\caption{\label{compare}Main differences between the ALS systems of KAGRA and LIGO.}
\begin{indented}
\item[]\begin{tabular}{@{}lll}
\br
&KAGRA&Advanced LIGO\\
\mr
From where to inject green&Central area&End stations\\
Optical fibers&Within the central area&Along the arms\\
Signals of the arm DoFs&Summations in electronics&Optical beat notes\\
Number of optical sensors&4&6\\
Number of SHG setups&2&3\\
\br
\end{tabular}
\end{indented}
\end{table}


The working principle of the KAGRA ALS can be summarized as follows. The two auxiliary lasers are phase-locked to the PSL with certain frequency offsets, which are on the order of tens of megahertz for ease of beat note detection. The frequency of the green laser coming from each SHG output is locked to the corresponding arm cavity by using the combination of a double-path acousto-optic modulator (AOM) and a voltage controlled oscillator (VCO) as a frequency actuator. At this point, each control signal applied to the corresponding VCO is equivalent to the difference between the PSL frequency and the resonant frequency of the corresponding arm cavity, on the assumption that the auxiliary laser is tightly locked to the PSL via the phase-locked loop (PLL). Therefore, the sum of the two control signals works as an error signal for CARM, while the difference of the two works as an error signal for DARM. Let us call such error signals ALS CARM and ALS DARM. By feeding them back to the PSL frequency and arm lengths, respectively, the main laser frequency is fixed to certain points with respect to the resonances of the arms.

A detailed optical and control diagram of the KAGRA ALS is shown in Figure \ref{alsconfig}. 
For the PLLs of the auxiliary lasers, beat notes between the PSL and the auxiliary laser are detected by photodetectors (PLL PD X and Y). For this purpose, we chose silicon photodiodes, which have fast response but relatively low responsivity for $1064\,\mathrm{nm}$ laser, because we can easily have a lot of optical power on the photodiodes. The signals of the beat notes are demodulated at phase-frequency discriminators by mixing them with local oscillator (LO) signals from radio frequency (RF) signal generators. One can change the frequency offset between the PSL and the auxiliary laser by changing the frequency of the LO. Each demodulated signal is processed by an analogue filter circuit and then fed back to the frequency of the auxiliary laser via the piezoelectric transducer of the laser, accompanied with the laser temperature control for slow drift compensation. The green laser from each auxiliary laser is frequency-shifted by a double path AOM.
Phase modulation for the Pound--Drever--Hall (PDH) technique \cite{Drever1983} is successively applied to it by an electro-optic modulator (EOM) in the laser room enclosure. Then the green lasers are injected to the X and Y arms from the back side of PR2 and SR2, respectively. The reflection from each arm cavity is picked off using a Faraday isolator and detected by a photodetector (PDH PD X or Y) to provide the PDH error signal. Each PDH error signal is filtered by another analogue circuit and fed back to a low-noise VCO that generates RF signals driving the AOM. The control signals, the voltages applied to the VCOs, are used to obtain the ALS CARM and ALS DARM error signals.
Because of the different control bandwidths, the CARM loop partially involves analogue control filters, while the DARM loop is entirely realized by a digital control.
The control signals are added with each other and then low-passed in analogue filters to obtain the ALS CARM error signal; the low-pass filtering makes the frequency response of the ALS CARM matched with the one for the main laser CARM signal (REFL CARM).
The ALS CARM and the REFL CARM are sent to the same CARM servo board so that the input signals for the servo can be gradually switched.
The servo has two output ports named slow and fast outputs,
which are fed back to the IMC length and the error point of the IMC control loop, respectively, to form a dual loop control \cite{Nagano2003}. Meanwhile, for the DARM control, the two control signals driving the VCOs are sampled and digitally processed by real-time controllers, and then fed back to the differential motion of ETMs. Once the ALS CARM and ALS DARM loops are fully engaged, the resonant frequency of each arm cavity can be tuned to a desired value with respect to the PSL frequency by changing the frequency of the LO for the corresponding PLL.

\subsection{Noise design of the ALS}\label{noisedesignsubsection}
Here the noise design of the KAGRA ALS system is described. 
The primary noise requirement is that the root mean square (RMS) frequency fluctuation of the main laser with respect to the resonant frequency of one arm cavity is smaller than the linewidth of the arm cavity. This ensures smooth handing over from the ALS system to the main laser signals with a help of the signal derived from the optical power in transmission as done in Advanced LIGO \cite{Staley2014}. In KAGRA, the designed value of the full width of an arm cavity is $33\,\mathrm{Hz}$ \cite{Aso2013}.

A more ambitious target can be given by the linewidth of the CARM cavity, which virtually serves as a single optical cavity corresponding to the CARM, with the full interferometer locked.
If the RMS frequency fluctuation is smaller than the CARM linewidth, the CARM control can be directly handed over from the ALS CARM to the REFL CARM. 
This will greatly simplify the arm locking process and contribute to increasing the duty cycle of the detector.
The full interferometer CARM linewidth is narrowed to $1.7\,\mathrm{Hz}$ for KAGRA \cite{Aso2013} due to the double cavity pole of the power recycling. 

\begin{figure}
\centering
\includegraphics[width=10cm,clip]{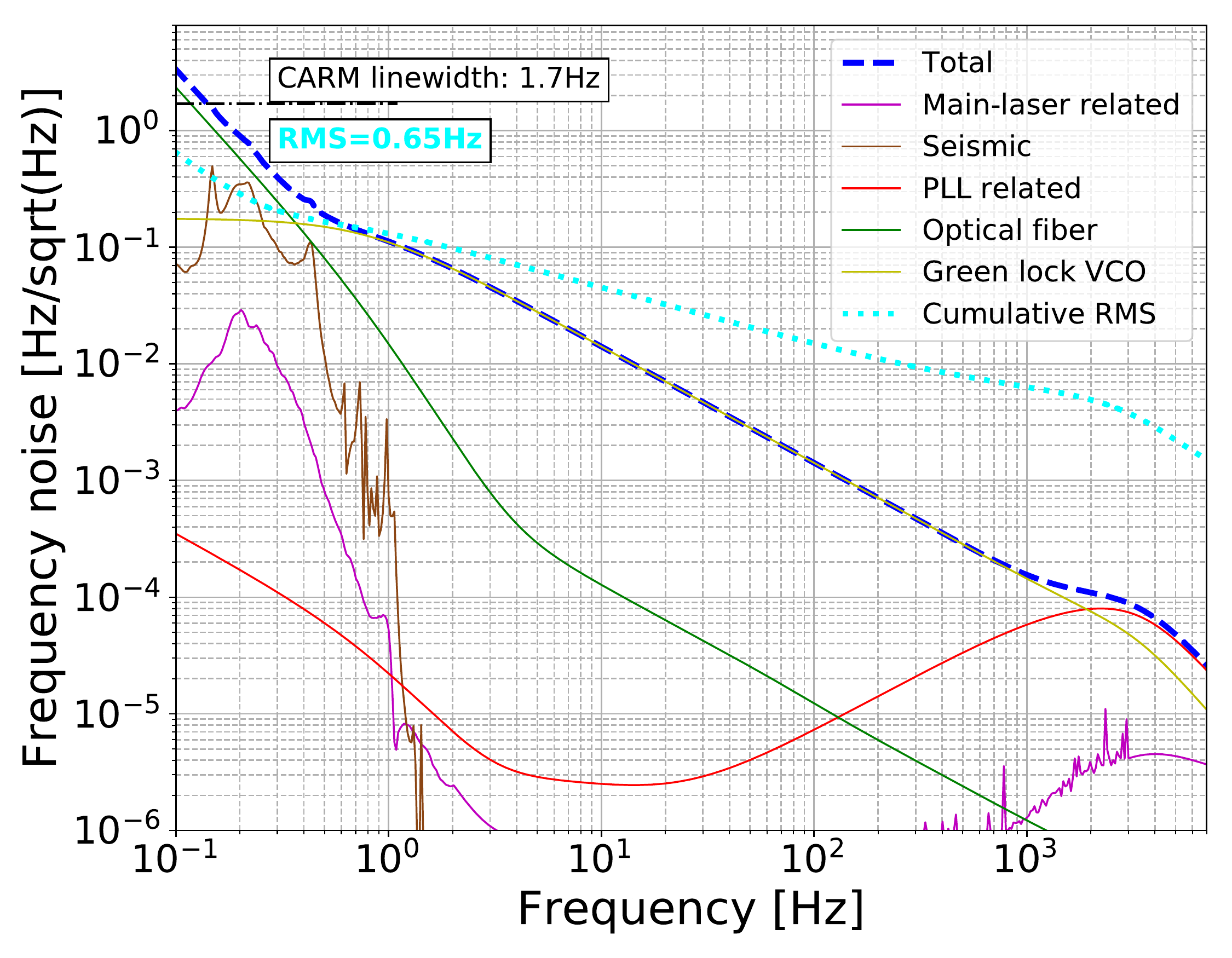}
\caption{Amplitude spectral densities of simulated noise of the KAGRA ALS system. The shown spectra are the noise contributions calibrated into $[\Delta f_\mathrm{main}(f)-\Delta f_\mathrm{CARM}(f)]L(f;\kappa_\mathrm{CARM})$. 
The blue dashed line shows total noise, which is quadrature sum of the contributions of all the noise sources considered in the design. The magenta trace shows the sum of the contributions of noises related to the PSL and its frequency stabilization. The brown line shows the estimated residual contribution of arm length fluctuations caused by seismic motion. The red trace shows the contribution of noises related to the PLLs of the auxiliary lasers including free-run frequency fluctuations of the auxiliary lasers. The green trace shows the contribution of phase noise caused by the optical fibers for the green lasers. The yellow trace shows the contribution of phase noises of the VCOs. The light blue dotted trace shows the cumulative RMS of the blue dashed trace integrated from the high frequency side.}\label{noisedesign}
\end{figure}
We numerically simulated and designed servo loops of the KAGRA ALS system in the frequency domain. The results of the simulated noise contribution to frequency fluctuations of the main laser with respect to the CARM resonant frequency are shown in Figure \ref{noisedesign}.
Since the PDH signal of a cavity senses low-passed fluctuations of the input laser frequency with respect to the cavity resonant frequency with the corner frequency of the low-pass equal to the cavity pole frequency \cite{Rakhmanov2002}, such low-passed fluctuations need to be considered for the evaluation of noise performance of the ALS system. 
Thus for the CARM control, the noise spectra have been computed by $[\Delta f_\mathrm{main}(f)-\Delta f_\mathrm{CARM}(f)]L(f;\kappa_\mathrm{CARM})$, where $\Delta f_\mathrm{main}$ and $\Delta f_\mathrm{CARM}$ are the frequency of the main laser entering the main interferometer and the CARM resonant frequency, respectively, and $L(f;\kappa_\mathrm{CARM})$ is a low-pass filter with a corner frequency of $\kappa_\mathrm{CARM}$, the cavity pole frequency of CARM \cite{Izumi2008}. $L(f;\kappa_\mathrm{c})$ is defined as
\begin{equation}
L(f;\kappa_\mathrm{c})= \frac{1}{1+if/\kappa_\mathrm{c}} .
\end{equation}
Here, $\Delta f_\mathrm{main}$ is the laser frequency fluctuation at the PRM, while $\Delta f_\mathrm{CARM}$ is the averaged fluctuation of the resonant frequencies of the X and Y arms.
We adopt the spectra of $[\Delta f_\mathrm{main}(f)-\Delta f_\mathrm{CARM}(f)]L(f;\kappa_\mathrm{CARM})$ because frequency fluctuations measured by the common mode of the arm cavities includes the effect from the CARM cavity pole and its RMS has to be compared with the CARM linewidth. 
Figure \ref{noisedesign} shows that the RMS of frequency fluctuations controlled by the ALS system is designed to be as low as $0.65\,\mathrm{Hz}$, which is smaller than the CARM linewidth. The dominant noise source above approximately $1\,\mathrm{Hz}$ is expected to be frequency noise of the VCOs driving the AOMs. The dominant noise source below $1\,\mathrm{Hz}$ is expected to be phase noise of the green lasers introduced by the optical fibers. Fiber phase noise had been estimated based on a measurement of phase noise of an optical fiber of length of $5\,\mathrm{m}$ placed on an optical table. We scaled the measured level of phase noise in such a way that the power spectral density of phase noise of a fiber is proportional to the length of the fiber. We also confirmed that the primary noise requirement, which is the target for the single arm, can be achieved; the RMS of $[\Delta f_\mathrm{main}(f)-\Delta f_\mathrm{ARM}(f)]L(f;\kappa_\mathrm{ARM})$ is $2.0\,\mathrm{Hz}$, which is smaller than the linewidth of an the arm cavity. Here, $\Delta f_\mathrm{ARM}$ is the resonant frequency of an arm cavity and $\kappa_\mathrm{ARM}$ is the cavity pole frequency of the arm cavity. The low-pass filter with the single arm pole $\Delta f_\mathrm{ARM}$ is used because this is the evaluation of the single arm performance of the ALS system.

\section{Experiment with the X arm cavity}
The performance of the KAGRA ALS system was evaluated with one of the arm cavities (the X arm).
The single-arm setup allowed us for testing the control system for CARM. This is equivalent to the frequency stabilization control using the arm cavity as a frequency reference. 
In this section, the parameters of the arm cavity measured during this experimental period are shown along with the design values after the setup of the experiment is overviewed. Subsequently, the lock acquisition process of the arm cavity we demonstrated is explained. Finally, the results of the detailed analysis of noises of the ALS system are shown. 

\subsection{Set-up}
\begin{figure}
\centering
\includegraphics[width=16cm,clip]{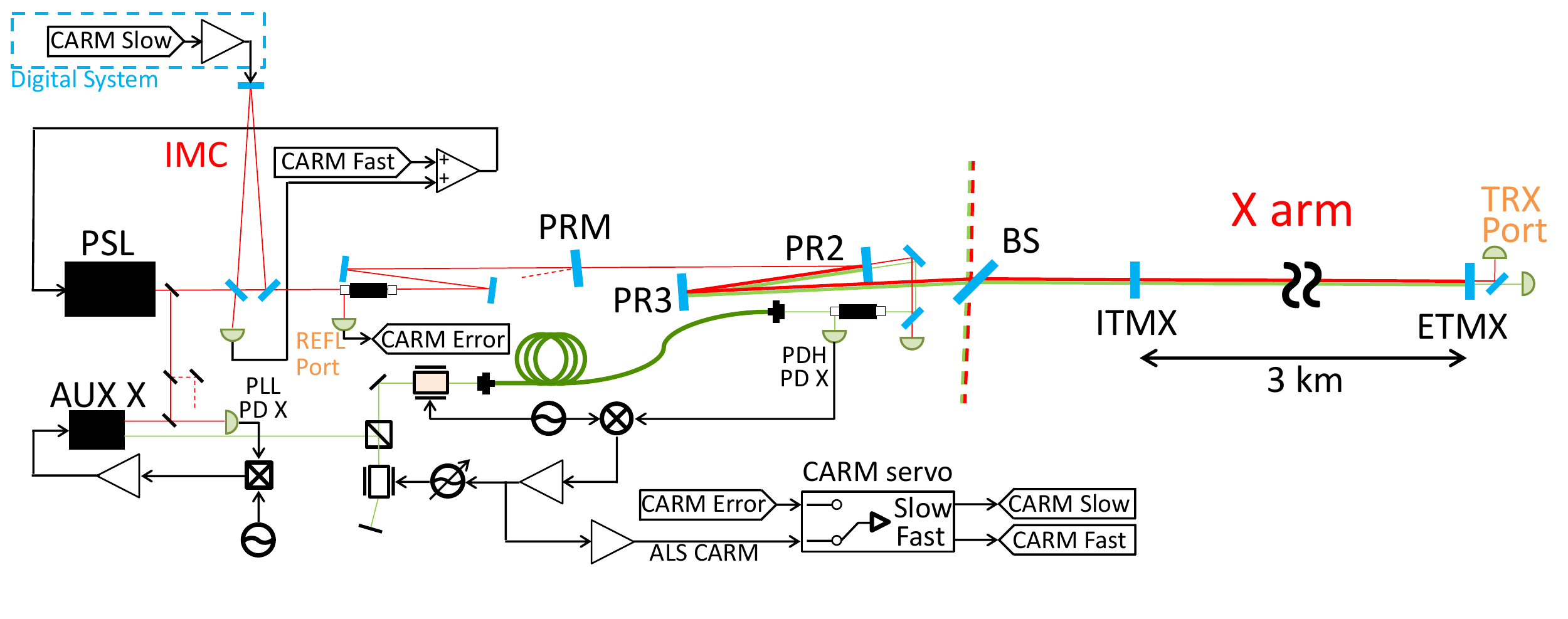}
\caption{Schematic view of the configuration of the interferometer and the control for the X arm lock experiment. The PRM was misaligned, and the Y arm and the signal recycling mirrors were not involved in the experiment. The legend for the symbols in the figure can be found in Figure \ref{alsconfig}.}\label{xarmconfig}
\end{figure}
Figure \ref{xarmconfig} shows the set-up for the length sensing and control of the X arm cavity during the experimental period. This experimental set-up was almost identical to the design shown in Figure \ref{alsconfig}. The differences were the following. Only the X arm was involved, and therefore the optics and electronics related to the Y arm were not involved. The PRM was intentionally misaligned. Unrelated laser beams were dumped so that they did not reach the signal recycling mirrors. 

The optical power of the main laser incident on the IMC was set to about $270\,\mathrm{mW}$. The laser power incident on the X arm was approximately $10\,\mathrm{mW}$, after the partial transmission of the PRM ($10\,\%$) and the BS ($50\,\%$). The main laser was phase-modulated at $45.0\,\mathrm{MHz}$ in the laser room. The power of the green laser after the optical fiber was approximately $10\,\mathrm{mW}$. The LO signal at about $40.0\,\mathrm{MHz}$ for the PLL was provided by a stable signal generator, E8663D from Keysight Technologies. The frequency of the oscillator for the PDH method of the X arm green laser was set to $33\,\mathrm{MHz}$. The center frequency of the VCO was approximately $80\,\mathrm{MHz}$. The PDH signal of the main laser and the X arm, which is labeled as ``CARM Error'' in Figure \ref{xarmconfig}, was obtained by the demodulation of the signal of the photodetector at the REFL port at $45.0\,\mathrm{MHz}$.

\subsection{Measured parameters of the X arm cavity}\label{sec_param}
\begin{table}
\caption{\label{parameters}Optical parameters of the X arm cavity. The values measured in the X arm experiment are shown along with the design values.}
\begin{indented}
\item[]\begin{tabular}{@{}lccl}
\br
Parameter name&Designed&Measured\\
\mr
Cavity length$^{a}$&$3000\,\mathrm{m}$  \cite{Aso2013}       &$2999.990(2)\,\mathrm{m}$\\
Finesse for $1064\,\mathrm{nm}$$^{b}$&$1530$ \cite{Aso2013}        &$1410(30)$\\
Roundtrip loss for $1064\,\mathrm{nm}$$^{a,c}$&$< 100\,\mathrm{ppm}$ \cite{Aso2013}        &$86(3)\,\mathrm{ppm}$\\
Mode matching ratio for $1064\,\mathrm{nm}$$^{a}$&--        &$0.91(1)$\\
Transverse mode spacing$^{a}$&$34.80\,\mathrm{kHz}$  \cite{Aso2013}       &$34.79(5)\,\mathrm{kHz}$\\
Finesse for $532\,\mathrm{nm}$$^{d}$&$49.2$        &$41.0(3)$\\
\br
\end{tabular}\\
Refer to the main body for the description of each measurement method. $^{a}$Measured by cavity scan; $^{b}$Measured by ring-down; $^{c}$Measured by the reflectivity of the cavity; $^{d}$Measured by the transfer function of the green lock loop.
\end{indented}
\end{table}
Table \ref{parameters} summarizes the measured parameters of the X arm cavity.
The cavity length, mode matching ratio of the main laser to the arm, and transverse mode spacing were measured by scanning the main laser frequency through free spectral ranges of the arm cavity utilizing the ALS system. The finesse for the main laser was measured by a ring-down method \cite{isogai2013}. The roundtrip loss of the arm cavity was obtained by the measurement of the reflectivity of the cavity combined with the information from the cavity scan measurement \cite{isogai2013}. The finesse for the green laser was obtained by the measurement of the open loop transfer function of the PDH lock loop. 

\subsection{Demonstration of the control handing-off}
\begin{figure}
\centering
\includegraphics[width=12cm,clip]{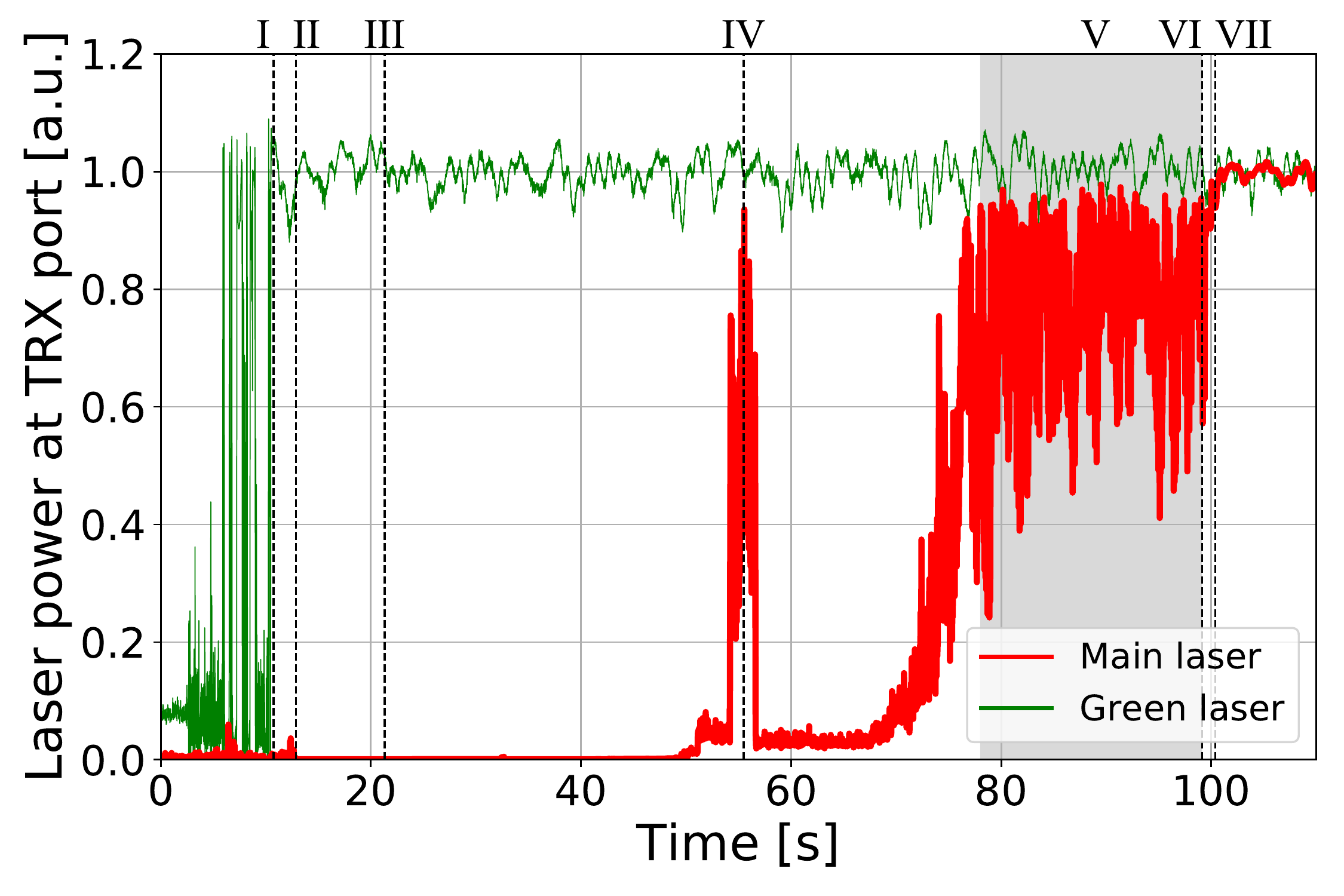}
\caption{Laser power inside the X arm cavity during the lock acquisition process. The two traces show the time series plots of the laser power detected by photodetectors at TRX port, the transmission monitor port for the X arm. The red trace corresponds to the power of the main laser inside the cavity, and the green trace correspond to the power of the green laser inside the cavity. The labels from I to VII show the timings when the steps of the lock acquisition sequence switched from one to another.}\label{demonstration}
\end{figure}
We achieved and demonstrated lock acquisition of the arm, where the error signal was handed over from the ALS CARM signal to the REFL CARM signal. Figure \ref{demonstration} shows how the power of the main and green laser at the TRX port (transmission of the X arm; depicted in Figure \ref{xarmconfig}) evolved in time, along with the steps taken throughout the entire process.
At point I, the resonance of the arm only for the green laser was achieved by locking the green laser frequency to the arm using the VCO only as an actuator. At point II, the feedback of the ALS CARM signal to the IMC length was turned on. From this point, the difference between the main laser frequency and the arm resonant frequency was controlled via the ALS system so that no longer the main laser stochastically resonated in the arm.
At point III, the ALS CARM control with the full frequency bandwidth was engaged using both slow and fast outputs of the CARM servo. The difference between the main laser frequency and the arm resonant frequency was gradually tuned by sweeping the LO frequency of the PLL. At point IV, the position of the resonance of the arm cavity was spotted in terms of the LO frequency. In the shaded area labeled as V, the main laser was kept at a resonant point via the control by the ALS system. We stayed for approximately $20\,\mathrm{s}$ in this stage for the purpose of demonstrating the stability of the ALS system. 
At point VI, the input of the CARM servo for ``CARM Error'' was enabled to start the hand-over of the control paths. At point VII, after $1\,\mathrm{s}$ from point VI, the input of the CARM servo from the ALS CARM signal was disabled to finish the hand-over. Consequently, lock acquisition of the main laser to the X arm was achieved. These lock acquisition processes did not fail unless the control of the green laser frequency to the arm lost its lock. Since the duration of lock of the green laser frequency to the arm was typically a few hours, it can be said that the processes were reliable. We tested these processes several times on different days, and succeeded in achieving lock acquisition of the main laser to the arm every time in the same way. This result clearly shows that the performance of the ALS system was demonstrated and the lock acquisition scheme of an arm cavity with it was established.

\subsection{Noise budget of the ALS of the X arm}
Here we present a characterization of the main noise sources of the ALS system.
To evaluate residual frequency fluctuations of the PSL with respect to the arm cavity, we locked the main laser frequency to the X arm with ``CARM Error'' used as an error signal while the frequency of the green laser injected to the X arm was also locked to the X arm without feeding any signals back to the main laser or test masses. In this configuration, sensing noise of the ALS system can be inferred by measuring the ALS CARM signal under the assumption that residual frequency fluctuations of the main laser with respect to the X arm is small enough to be ignored for the noise estimation,
which should be validated.

\begin{figure}
\centering
\includegraphics[width=12cm,clip]{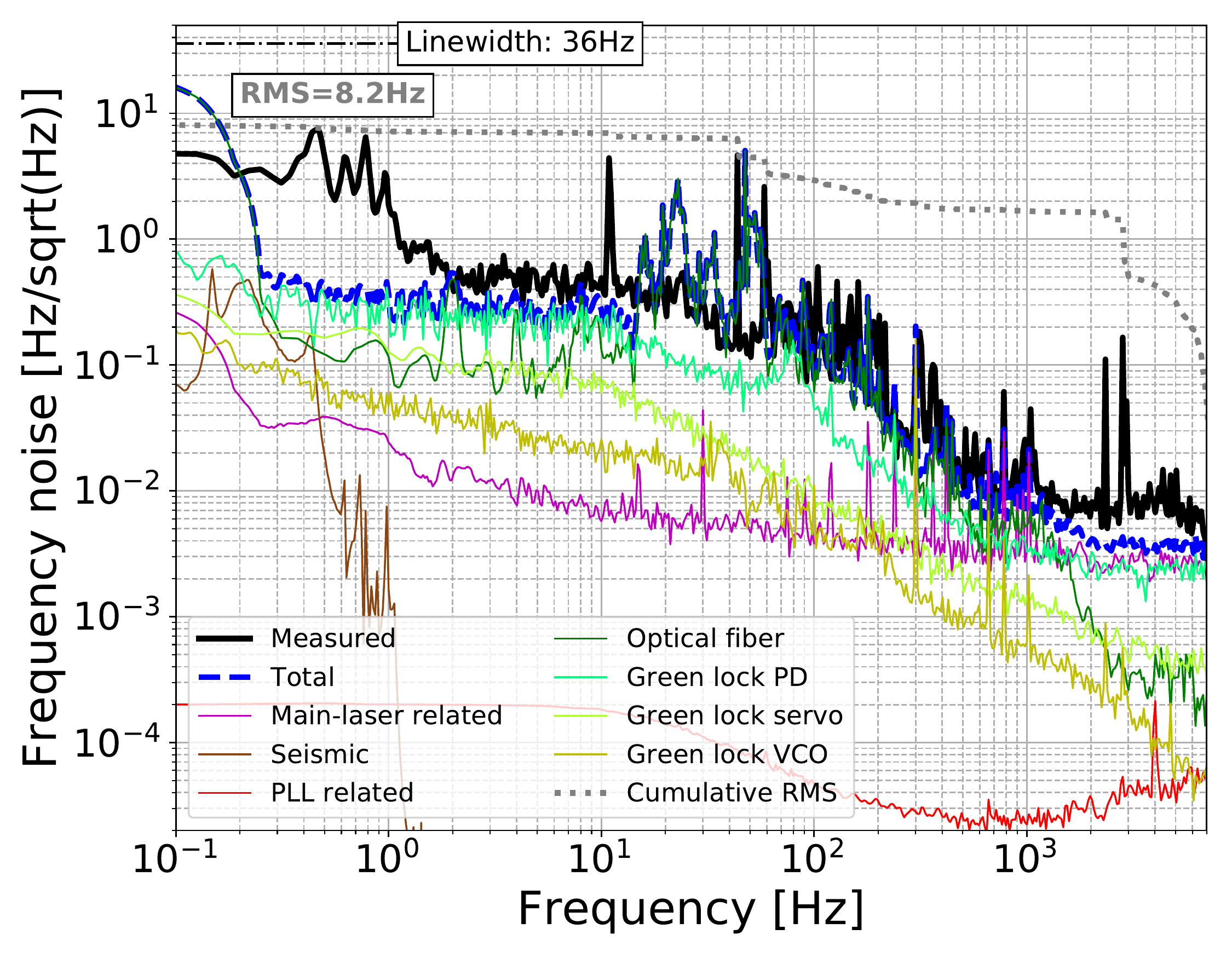}
\caption{Noise budget plot of sensing noise of the KAGRA ALS. 
We evaluated the sensing noise instead of directly measuring the ASD of ``CARM Error'' with the PSL frequency controlled via the ALS system, in order to eliminate possible nonlinear behavior in the ``CARM Error'' PDH signal.
The black trace shows the measured ASD of sensing noise of the ALS, calibrated to $[\Delta f_\mathrm{main}(f)-\Delta f_\mathrm{X}(f)]L(f;\kappa_\mathrm{X})$. The grey dotted trace shows the cumulative RMS of the black trace from the high frequency side. The blue dashed line shows total noise, quadrature sum of the contributions of all the known noise sources described in the following. The magenta trace shows the sum of the contributions of noises related to the PSL and its frequency stabilization. The brown line shows the estimated residual contribution of arm length fluctuations caused by seismic motion. The red trace shows the contribution of noises associated with the PLL of the auxiliary laser including free-run frequency fluctuations of the auxiliary laser. The green trace shows the contribution of phase noise caused by the optical fiber for the green laser. The light green trace shows the contribution of noise of the photodetector circuit to obtain the PDH signal of the green laser. The yellow-green trace shows the contribution of noise of the servo circuit for the feedback to the VCO. The yellow trace shows the contribution of phase noise of the VCO.}\label{noisebudget}
\end{figure}
The results of the ALS sensing noise are shown in Figure \ref{noisebudget}. The black trace shows the measured amplitude spectral density (ASD) of the ALS sensing noise calibrated to $[\Delta f_\mathrm{main}(f)-\Delta f_\mathrm{X}(f)]L(f;\kappa_\mathrm{X})$, where $\Delta f_\mathrm{X}$ is the resonant frequency of an arm cavity and $\kappa_\mathrm{X}$ is the cavity pole frequency of the arm cavity. The grey dotted trace shows the cumulative RMS of the black trace integrated from the high frequency side. The RMS was measured to be $8.2\,\mathrm{Hz}$, which is smaller than the primary requirement, i.e., the linewidth of the X arm cavity. This indicates that the ALS system was sensitive enough to hold the main laser within the resonance width of one arm cavity, as is consistent with the time series shown in the previous section. Note that here we used the measured parameters of the cavity described in Section \ref{sec_param}. 

In order to investigate the origin of measured sensing noise of the ALS system, 
we made a number of supplemental measurements to estimate the contributions of the various noise sources. Figure \ref{noisebudget} shows the results.
The blue dashed line shows total noise, which is quadrature sum of the contributions of all the considered noise sources. In the frequency band approximately between $10\,\mathrm{Hz}$ and $1000\,\mathrm{Hz}$, phase noise caused by the optical fiber for the green laser limited total noise. It shows a good agreement with measured sensing noise. In contrast, there is discrepancy between total and measured noise at low frequencies below $10\,\mathrm{Hz}$. This indicates there were noise sources that are not taken into account.
The magenta trace includes the residual fluctuations of the main laser frequency and was lower than the black trace, which validated the assumption of this noise evaluation scheme.

\section{Discussion}
\subsection{Noise sources}
Let us discuss here the implications of the noise analysis shown in Figure \ref{noisebudget}. 
Comparing Figures \ref{noisedesign} and \ref{noisebudget}, it is clear that we had underestimated the fiber phase noise. 
This probably came from the difference of the environment between the KAGRA site and the fiber testing setup. The optical fibers in KAGRA were laid on cable racks and the wall of the laser room enclosure, and they suffer from vibration and air flow.
Since it is known that phase noise caused by an optical fiber can be largely suppressed by a fiber noise cancellation technique \cite{ye2003}, we can suppress this noise by implementing such a technique if further improvement of the noise performance is required. 

\begin{figure}
\centering
\includegraphics[width=10cm,clip]{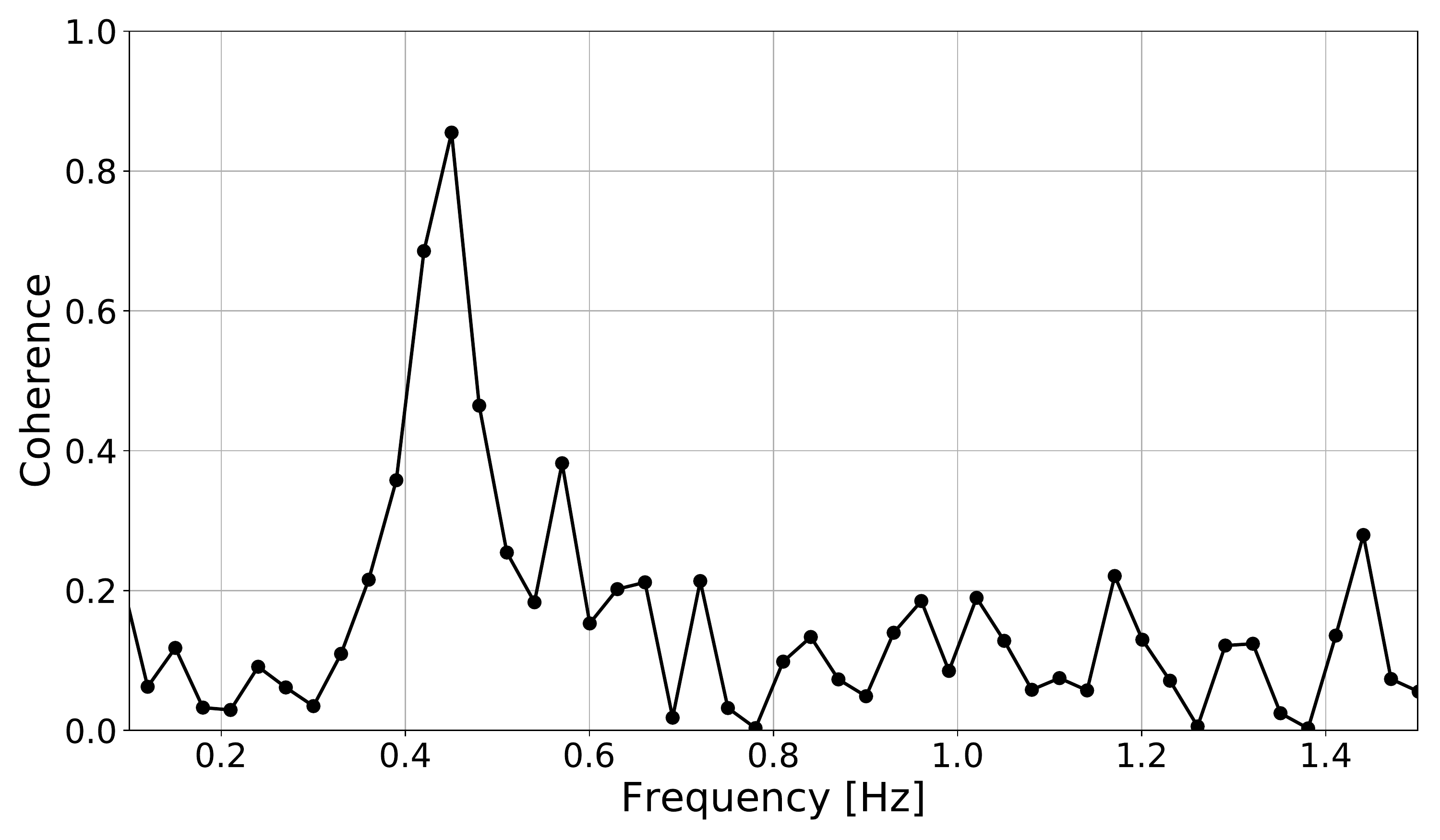}
\caption{Spectrum of the coherence between the longitudinal motion of PR2 and sensing noise of the ALS system.}\label{PR2_coh}
\end{figure}
The possible noise sources that might account for the discrepancy between total and measured noise at low frequencies below $10\,\mathrm{Hz}$ are the couplings from the motion of the suspended optics. We observed significant coherence between measured sensing noise of the ALS system and the longitudinal motion of PR2 sensed by its local sensors \cite{akiyama2019} at the peak at $0.45\,\mathrm{Hz}$ (Figure \ref{PR2_coh}). PR2 reflects the main laser but transmits the green laser. Therefore, its longitudinal motion must have caused Doppler shift only in the frequency of the main laser. 
The amount of the PR2 motion sensed by the local sensors was $0.85\,\mathrm{\mu m/\sqrt{Hz}}$ at $0.45\,\mathrm{Hz}$. Since the amount of the Doppler shift can be expressed as $2v/\lambda$, where $v$ is velocity of the mirror and $\lambda$ is wavelength of the main laser, the Doppler shift caused by PR2 is estimated to be $4.5\,\mathrm{Hz/\sqrt{Hz}}$, which is comparable to the peak height in the sensing noise at $0.45\,\mathrm{Hz}$.
On this Doppler noise issue, one strategy for improvement is an online noise subtraction utilizing the local sensors.
Similarly, the motion of the steering mirror behind PR2 that is installed on a suspended breadboard might have introduced Doppler noise because this mirror is relevant only to the green laser. 
Although there was no direct way to prove it because we had no sensors for this mirror unfortunately, it is not likely that the Doppler shift caused by the steering mirror limited the noise performance; it can be inferred that the contribution of such a Doppler shift was smaller the sensing noise level of the ALS system, on the assumption that the amount of the motion of the steering mirror is on the same order of the seismic motion.
In any case, we need further investigation on the discrepancy.

\subsection{Towards lock of the full interferometer}
\begin{figure}
\centering
\includegraphics[width=10cm,clip]{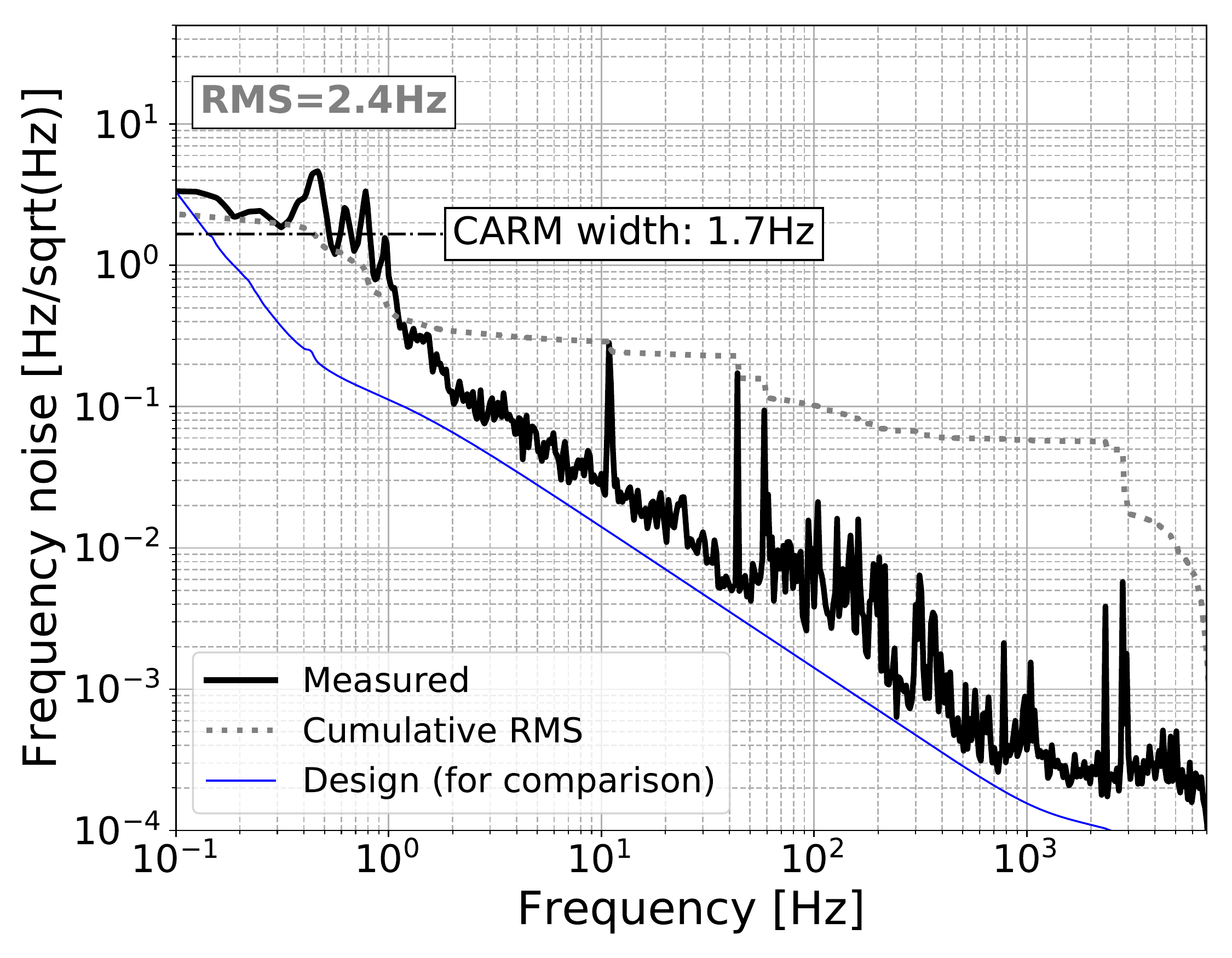}
\caption{Expected spectrum of the ASD of $[\Delta f_\mathrm{main}(f)-\Delta f_\mathrm{CARM}(f)]L(f;\kappa_\mathrm{CARM})$, calculated from measured sensing noise of the ALS of the X arm. We assumed that sensing noises of the ALS of the X and Y arm are independent. The grey dotted trace shows the cumulative RMS of the black trace from the high frequency side. The designed noise level of the ALS system are shown again by the blue trace for comparison.}\label{alscarmexpect}
\end{figure}

Next, we discuss the demonstrated performance of the ALS system in the context of  the full interferometer of KAGRA. As mentioned in Section \ref{noisedesignsubsection}, 
the measured RMS of the ALS sensing noise was well within the primary requirement ($33\,\mathrm{Hz}$).
Therefore, assuming that we can achieve the same noise performance of the ALS system also for the Y arm, we will be able to lock the full interferometer using the ALS system with a help of the arm transmission signals \cite{Staley2014}. Keeping the same assumption, we expect that the RMS of $[\Delta f_\mathrm{main}(f)-\Delta f_\mathrm{CARM}(f)]L(f;\kappa_\mathrm{CARM})$ will be $2.4\,\mathrm{Hz}$ (Figure \ref{alscarmexpect}). Figure \ref{alscarmexpect} indicates that we need to improve the noise level of the ALS system in the frequency band lower than approximately $1\,\mathrm{Hz}$ to satisfy the more ambitious target, which is that the RMS of $[\Delta f_\mathrm{main}(f)-\Delta f_\mathrm{CARM}(f)]L(f;\kappa_\mathrm{CARM})$ is smaller than the full width of the CARM cavity. Although the current noise performance does not satisfy the more ambitious target, a previous work \cite{izumi2014a} suggested that a self-amplification process of the PDH signal might enable us to hand over the control paths directly from the ALS CARM to the REFL CARM signal, even if frequency fluctuations are larger than the linewidth. Achieving direct lock of CARM with the ALS system for the first time is a future work.

\section{Conclusion}
We designed a new type of the ALS system for KAGRA. 
It is scalable to the GW detectors even with longer arms such as the ones planned in the third generation detectors, thanks to the vertex injection of the green lasers to the arms.
The detailed noise design of the ALS showed that the ALS can potentially enable us to directly hand over the CARM control from the ALS CARM to the PDH signal of the main laser of CARM, which will simplify the lock acquisition process. 
Utilizing the X arm cavity, the performance of the new ALS was tested and characterized as follows.
(1) Lock acquisition of the arm cavity through the hand-over from the ALS CARM to the PDH signal of the main laser were demonstrated.
(2) The noise characterization of the system showed that the RMS of $[\Delta f_\mathrm{main}(f)-\Delta f_\mathrm{X}(f)]L(f;\kappa_\mathrm{X})$ reached $8.2\,\mathrm{Hz}$, with $\Delta f_\mathrm{main}(f)$ controlled by the ALS system. The RMS was well within the linewidth of the single arm cavities ($33\,\mathrm{Hz}$) and thus the noise performance should be sufficient for lock acquisition of the full interferometer of KAGRA.
These results show that the new ALS system has been experimentally established. 

\ack
This work was supported by MEXT, JSPS Leading-edge Research Infrastructure Program, JSPS Grant-in-Aid for Specially Promoted Research 26000005, MEXT Grant-in-Aid for Scientific Research on Innovative Areas 24103005, JSPS Core-to-Core Program, A. Advanced Research Networks, the joint research program of the Institute for Cosmic Ray Research, University of Tokyo, National Research Foundation (NRF) and Computing Infrastructure Project of KISTI-GSDC in Korea, the LIGO project, and the Virgo project.

\section*{References}

\begin{thebibliography}{10}

\bibitem{Abbott2016}
B.~P. Abbott et~al.,
\newblock Phys. Rev. Lett. {\bf 116}, 061102 (2016).

\bibitem{Abbott2017a}
B.~P. Abbott et~al.,
\newblock Phys. Rev. Lett. {\bf 119}, 161101 (2017).

\bibitem{Abbott2017b}
B.~P. Abbott et~al.,
\newblock Astrophys. J. Lett. {\bf 848}, L12 (2017).

\bibitem{Scenario2018}
{KAGRA Collaboration, LIGO Scientific Collaboration and Virgo Collaboration}
  et~al.,
\newblock Living Rev. Relativ. {\bf 21}, 3 (2018).

\bibitem{wen2010}
L.~Wen and Y.~Chen,
\newblock Phys. Rev. D {\bf 81}, 082001 (2010).

\bibitem{takeda2018}
H.~Takeda, A.~Nishizawa, Y.~Michimura, K.~Nagano, K.~Komori, M.~Ando, and
  K.~Hayama,
\newblock Phys. Rev. D {\bf 98}, 022008 (2018).

\bibitem{Akutsu2018}
T.~Akutsu et~al.,
\newblock Prog. Theor. Exp. Phys. {\bf 2018}, 013F01 (2018).

\bibitem{Somiya2012}
K.~Somiya,
\newblock Class. Quantum Grav. {\bf 29}, 124007 (2012).

\bibitem{akutsu2019}
T.~Akutsu et~al.,
\newblock Class. Quantum Grav. {\bf 36}, 165008 (2019).

\bibitem{Hild2011}
S.~Hild et~al.,
\newblock Class. Quantum Grav. {\bf 28}, 094013 (2011).

\bibitem{abbott2017}
B.~P. Abbott,
\newblock Phys. Rev. Lett. {\bf 118}, 221101 (2017).

\bibitem{acernese2006}
F.~Acernese et~al.,
\newblock Class. Quantum Grav. {\bf 23}, S85 (2006).

\bibitem{evans2002}
M.~Evans et~al.,
\newblock Opt. Lett. {\bf 27}, 598 (2002).

\bibitem{mullavey2012}
A.~J. Mullavey, B.~J.~J. Slagmolen, J.~Miller, M.~Evans, P.~Fritschel, D.~Sigg,
  S.~J. Waldman, D.~A. Shaddock, and D.~E. McClelland,
\newblock Opt. Express {\bf 20}, 81 (2012).

\bibitem{Izumi2012}
K.~Izumi et~al.,
\newblock J. Opt. Soc. Am. A {\bf 29}, 2092 (2012).

\bibitem{Staley2014}
A.~Staley et~al.,
\newblock Class. Quantum Grav. {\bf 31}, 245010 (2014).

\bibitem{Acernese2015}
F.~Acernese et~al.,
\newblock Class. Quantum Grav. {\bf 32}, 024001 (2015).

\bibitem{yeaton-massey2012}
D.~Yeaton-Massey and R. X Adhikari,
\newblock Opt. Express {\bf 20}, 19 (2012).

\bibitem{drever1991}
R. W. P. ~Drever,
\newblock {\it The Detection of Gravitational Wave}, edited by D. G. ~Blair, Cambridge University Press (1991).

\bibitem{meers1988}
B. J.~Meers,
\newblock Phys. Rev. D {\bf 38}, 2317-26 (1988).

\bibitem{mizuno1993}
J.~Mizuno et~al.,
\newblock Phys. Lett. A {\bf 175}, 273 (1993).

\bibitem{fricke2012}
T.T.~Fricke et~al.,
\newblock Class. Quantum Grav. {\bf 29}, 065005 (2012).

\bibitem{regehr1995}
M.W.~Regehr, F.J.~ Raab, and S.E.~Whitcomb,
\newblock Opt. Lett. {\bf 20}, 1507 (1995).

\bibitem{martynov2016}
D. V.~Martynov et~al.,
\newblock Phys. Rev. D {\bf 93}, 112004 (2016).

\bibitem{Aso2013}
Y.~Aso, Y.~Michimura, K.~Somiya, M.~Ando, O.~Miyakawa, T.~Sekiguchi,
  D.~Tatsumi, and H.~Yamamoto,
\newblock Phys. Rev. D {\bf 88}, 043007 (2013).

\bibitem{arai2000}
K.~Arai, M.~Ando, S.~Moriwaki, K.~Kawabe, and K.~Tsubono,
\newblock Phys. Lett. A {\bf 273}, 15 (2000).

\bibitem{Yamamoto2019}
K.~Yamamoto et~al.,
\newblock Accepted by Class. Quantum Grav. (2019), https://doi.org/10.1088/1361-6382/ab4489.

\bibitem{ye2003}
J.~Ye et~al.,
\newblock J. Opt. Soc. Am. B {\bf 20}, 1459 (2003).

\bibitem{Drever1983}
R.~W.~P. Drever, J.~L. Hall, F.~V. Kowalski, J.~Hough, G.~M. Ford, A.~J.
  Munley, and H.~Ward,
\newblock Appl. Phys. B {\bf 31}, 97
  (1983).

\bibitem{Nagano2003}
S.~Nagano et~al.,
\newblock Rev. Sci. Instrum. {\bf 74}, 4176 (2003).

\bibitem{Rakhmanov2002}
M.~Rakhmanov, R.L.~Savage Jr., D.H.~Reitze, and D.B.~Tanner,
\newblock Phys. Lett. A {\bf 305}, 239-244 (2002).

\bibitem{Izumi2008}
K.~Izumi and D.~Sigg,
\newblock Class. Quantum Grav. {\bf 34}, 015001 (2017).

\bibitem{isogai2013}
T.~Isogai, J.~Miller, P.~Kwee, L.~Barsotti, and M.~Evans,
\newblock Opt. Express {\bf 21}, 30114 (2013).

\bibitem{akiyama2019}
Y.~Akiyama et~al.,
\newblock Class. Quantum Grav. {\bf 36}, 095015 (2019).

\bibitem{izumi2014a}
K.~Izumi, D.~Sigg, and L.~Barsotti,
\newblock Opt. Lett. {\bf 39}, 5285 (2014).

\end{thebibliography}

\end{document}